# Punishing defectors and rewarding cooperators: Do people discriminate between genders?

Valerio Capraro[1] and Hélène Barcelo[2]

**Abstract**

Do people discriminate between men and women when they have the option to punish defectors or reward cooperators? Here we report on four pre-registered experiments, that shed some light on this question. Study 1 (N=544) shows that people do not discriminate between genders when they have the option to punish (reward) defectors (cooperators) in a one-shot prisoner's dilemma with third-party punishment/reward. Study 2 (N=253) extends Study 1 to a different method of punishing/rewarding: participants are asked to rate the behaviour of a defector/cooperator on a scale of 1 to 5 stars. In this case too, we find that people do not discriminate between genders. Study 3a (N=331) and Study 3b (N=310) conceptually replicate Study 2 with a slightly different gender manipulation. These latter studies show that, in situations where they do not have specific beliefs about the gender of the defector/cooperator's partner, neither men nor women discriminate between genders.

*Keywords:* punishment, reward, cooperation, gender.

*JEL Classification: D02, D03, D63, D69, D79, H41*

---

[1] Economics Department, Middlesex University London, UK. Contact author: v.capraro@mdx.ac.uk
[2] Mathematical Science Research Institute, Berkeley, CA, USA. This author is supported by the US National Science Foundation under the grant No. DMS-1440140.

# 1. Introduction

Do people discriminate between men and women when they have the option to punish defectors or reward cooperators? Answering this question has potentially far-reaching implications on gender equity, since cooperative behaviour forms the basis of our societies and is typically enforced through punishment or rewarding (Andreoni, Harbaugh, & Vesterlund, 2003; Boyd & Richerson, 1992; Fehr & Gächter, 2002; Gürerk, Irlenbusch, & Rockenbach, 2006; Hilbe & Sigmund, 2010; Milinski, Semmann, Krambeck, & Marotzke, 2006; Nowak, 2006; Perc et al. 2017).

A handful of works have used stylised economic games, such as the ultimatum game and the public goods game, to explore the "mirror question", whether men and women punish differently (Eckel & Grossman, 2001; Solnick, 2001; García-Gallego, Georgantzís, & Jaramillo-Gutiérrez, 2012; Bueno-Guerra, Leiva, Colell, & Call, 2016; Burnham, 2018). Although important in their domain, these studies do not allow us to answer our research question. To the best of our knowledge, only one study explored if men and women *are punished* differently using economic games. Gürerk, Irlenbusch, and Rockenbach (2018) conducted a repeated public goods game in which participants, at each round, can decide whether to play a public goods game in a mixed-sex group or in a same-sex group. Each iteration was followed by a punishment phase. Gürerk et al. found that the average number of punishment points received by women is not statistically different from the average number of punishment points received by men. This result thus supports the hypothesis:

*H1. People do not discriminate between men and women when they have the option to punish defectors or reward cooperators.*

Gürerk et al. focused on a situation in which punishment can be used strategically to induce cooperation in the next round. Indeed, while groups change over time, there is a non-zero probability of playing again with the same participant, especially when participants punish same-gender partners. To the best of our knowledge, no studies have explored our research question in the context of one-shot cooperation games with punishment or reward.

Beyond cooperation games, some works have explored whether women and men are penalized or rewarded differently for the same behaviour in the workplace. Rudman (1998) found that self-promotion among women leads them to receive higher competence ratings, but it makes them perceived to be less socially attractive and hireable. Heilman and Chen (2005) found that women are penalized when they fail to act altruistically in the workplace, and they are not rewarded when they act altruistically; by contrast, men are rewarded when they act altruistically, and they are not penalized when they fail to act altruistically. Heilman and Chen also showed that this effect is driven by different prescriptive stereotypes. Bowles, Babcock, and Lai (2007) found that evaluators penalise women more than men for initiating negotiations. Moreover, women are less inclined to negotiate with male evaluators. No gender differences were found when the evaluator was female. Babcock, Recalde, Vesterlund, and Weingart (2017) found that women are more likely than men to volunteer, asked to volunteer, and accept requests to volunteer for low promotability tasks. And this was partly driven by the beliefs that women are more likely than men to say yes to low promotability tasks. Several studies have shown that students rate female instructors less than their male counterparts (MacNell, 2015; Boring, 2017; Mengel, Sauermann, & Zölitz, 2019). These works thus suggest that women are penalised more than men, especially in contexts where they are expected to behave in a communal way. We are not aware of any study testing whether women are expected to cooperate more than men in economic games, but there are two meta-analyses showing that women do cooperate more than men in prisoner's dilemma and public goods games (Balliet et al. 2011, Rand, 2017). Assuming that people have correct beliefs about the cooperative behaviour of men and women, we can then formulate the following alternative hypothesis.

*H2. Women are punished (rewarded) more (less) than men when they defect (cooperate).*

In sum, the question whether people discriminate between men and women when they have the option to punish defectors or reward cooperators remains largely unanswered and at least two alternative hypotheses can be supported based on previous related literature. In this paper, we move a step towards answering this question using a prisoner's dilemma with third-party punishment (reward) and through four pre-registered studies and a pre-study. The reason we chose the prisoner's dilemma is that it is the simplest game that captures the tension between cooperation and self-interest, having only two players, each of whom has only two strategies. The reason we chose third-party punishment/reward, instead of peer-punishment/reward is twofold. First, we wanted to formalise a situation that might be relevant in real life. For example, sometimes in a school setting, students are paired to work on a common project. Are the students punished or rewarded differently by teachers depending on their gender when they defect or cooperate? Another example is a defendant in a bench trial, where only one person, a judge, decides on the faith of the defendant. Second, as mentioned above, there is already one study exploring whether women and men are punished differently by peers (Gürerk et al. 2018), whereas, to the best of our knowledge, there are no studies exploring this question from the point of view of a third-party.

## 2. Study 1

In a pre-study (see Supplementary Material, Part 1), we collected the decisions of participants playing a prisoner's dilemma (PD) knowing that a third-party will be informed about their decision and will have the possibility to punish or reward them. Here, in Study 1, we collect the punishing and rewarding decisions. Since our goal is to explore whether women and men are punished (rewarded) differently when they defect (cooperate) in the prisoner's dilemma, we implement a 2 (punish, reward) x 2 (female, male) between-subjects design.

### 2.1. Methods

The experiment was conducted on Amazon Mechanical Turk (AMT; Paolacci, Chandler, & Ipeirotis, 2010). Participants were randomly divided in four conditions, the *Punish-Anna* condition, the *Punish-Adam* condition, the *Reward-Anna* condition, and the *Reward-Adam* condition. These four conditions were implemented similarly, and from here on, in order to simplify the description of the method, we will only describe the *Punish-Anna* condition.

In the *Punish-Anna* condition, participants played in the role of Participant C. They were shown the instructions of a PD, in which two players, named Anna and Participant B, make their decision knowing that they might be punished by Participant C. After reading the instructions of the PD, Participant C was given 40 points and was told that they could use these points to deduct points from the account of Anna, Participant B, or both. For every deduction point used, the account of Participant C would be decreased by 1 point, while the account of Anna or Participant B would be decreased by 3 points. To avoid negative payoffs, the number of points that could be actually used to deduct points was capped depending on

Anna's and Participant B's decisions. Participant C was informed that 1 point would correspond to 1 US cent. After reading the instructions, Participant C was asked seven comprehension questions to make sure they understood the game. Participants C failing at least one comprehension question were automatically excluded from the survey and received no payment; this exclusion criterium was described at the beginning of the survey. To make sure that those participants had also understood the gender of Anna, some of the questions contained the name Anna and the gendered pronoun her. For example, one comprehension question was: "In Stage 1, what is the choice that Anna should make in order to maximize her number of points?" (We also asked the participants of the additional studies 3a-3b which gender they thought Anna was, and 94% responded that Anna was a woman). The Participants C that passed the comprehension questions were requested to make their punishing decisions in eight possible scenarios, one of which would be randomly selected to determine their payment. The eight scenarios corresponded to all four possible combinations of Anna's and Participant B's choices regarding cooperating or not, times two, corresponding to the fact that participants were asked to deduct (or not) points from Anna's and Participant B's pools. After making their decisions, Participants C entered a demographic questionnaire, at the end of which they received the completion code needed to claim their payment through AMT. Full experimental instructions are reported in the Supplementary Material, Part 5.

Once the experiment ended, the bonuses were computed in such a way that all decisions were incentive compatible (see Supplementary Material, Part 2). Participants received their bonus in addition to the participation fee (50 cents). The design, the exclusion criteria, and the analysis were pre-registered at: http://aspredicted.org/blind.php?x=tc76r2. In line with hypothesis H2 in the Introduction, we pre-registered that: (i) women are punished more than men when they defect; and (ii) women are rewarded less than men when they cooperate.

### 2.2. Results

As pre-registered, whenever we found more than one observation with the same IP address or Turk ID, we kept only the first observation, as determined by the starting date, and discarded the rest.[3] In doing so, we were left with a total of N=544 participants (52% females; mean age = 38.2).[4]

### 2.2.1. *Punishment*

Figure 1 reports the number of points used by Participants C to punish Anna (Adam) and Participant B[5], split by the prisoner's dilemma choices made by Anna (Adam) and Participant B. In the x-axis of the figure, the first component of each pair corresponds to the contributions made by the participant to be punished. For example, the first light grey column represents the average number of points used to punish Adam when he cooperates while the other participant, that is, Participant B, defects. Similarly, the last dark grey column reports the average number of points used to punish Participant B when s/he defects while the other participant, which can be either Anna or Adam, cooperates. We clearly note that the extent to which a defector is punished strongly depends on the choice of the other player. Specifically, the average number of points used to punish a defector when the defector's partner *cooperates* is 2.16,[6] which is significantly higher than the average number of points used to punish a defector when the defector's partner *defects*, which is 0.38 points (linear regression[7]: $b = 1.781$, $t = 13.21$, $p < .001$).[8] In turn, the average number of points used to punish a defector when the defector's partner defects is significantly greater than the average number of points used to punish a cooperator[9], both when the cooperator's partner cooperates (0.22 points used to punish, on average; comparison vs. punishing a defector when their partner defects: $b = 0.160$, $t=2.32$, $p=0.020$) and when their partner defects (0.18 points used to punish, on average; comparisons vs. punishing a defector when their partner defects: $b = 0.196$, $t=3.31$, $p=0.001$).

---

[3] This is a standard procedure to increase the quality of data gathered using Amazon Mechanical Turk (Horton, Rand, & Zeckhauser, 2011; Berinsky, Margolis, & Sances, 2014; Litman & Robinson, 2020).

[4] The original sample size was N = 2,030. A total of 73% of the observations were excluded due to lack of comprehension or repeated Turk IDs or repeated IP addresses. The vast majority of these subjects were excluded due to lack of comprehension. On AMT studies using prisoner's dilemma games, it is common to lose about 10% of the sample for each comprehension question (Capraro, Jordan, & Rand, 2014). In our case, we had seven comprehension questions, to make sure that participants who pass them have understood all the key points of the game.

[5] Notice that there are two groups of subjects that are asked to punish Participant B, those in the Punish-Anna condition and those in the Punish-Adam condition. However, there is no significant difference on how these subjects punish Participant B (all p's > 0.273). Therefore, in the analysis we collapse the data of these two groups. Similarly, in the next subsection, we will collapse the data regarding rewarding Participant B.

Coming to gender differences, we use linear regression to compare the number of points used to punish Adam/Anna with and without control on sex, age, and education (of the punisher), and with and without adding the interaction between condition (*Punish-Anna* vs *Punish-Adam*) and gender of the punisher. Regression tables are reported in the Supplementary Material, Tables A2-A3. In short, in contrast to our pre-registered hypothesis, we find no evidence that Adam and Anna are punished differently and no evidence that men and women punish Adam and Anna differently.

---

[6] Note that the maximum number of punishment points that can be used to punish a defector when the other person cooperates is 11 (see Supplementary Material, Part 1, for details of the pre-study). Therefore, 2.16 points corresponds to 19.6% of the maximum permitted punishment, which is in line with previous literature. For example, Fehr & Fischbacher (2004) report an average punishment of 14% in a dictator game with third-party punishment, whereas Gürerk et al. (2018) report an average punishment of 10.8% in a public goods game with second-party punishment. Looking at the distribution of punishment choices, we found that 82% do not punish, 3% use 1 point to punish, 3% use 2 points, 2% use 3 points, 4% use 5 points, 2% use 6 points, 1% use 7 points, and the remaining 1% is spread among the remaining choices.
[7] All the results presented in this paragraph and in the corresponding paragraph of the Rewarding section are robust to using rank-sum test.
[8] This also implies that cooperation pays if one expects the partner to cooperate (one is punished by a little under 2 points less and hence saves nearly 6 cents, which exceeds the 5 cents saved by defecting). However, if a participant has correct expectations about the likelihood of being matched with a cooperator, punishment is not large enough to make cooperation pay.
[9] In line with previous work on anti-social punishment (Herrmann, Thöni, & Gächter, 2008), we also find that a small proportion of participants punish cooperators.

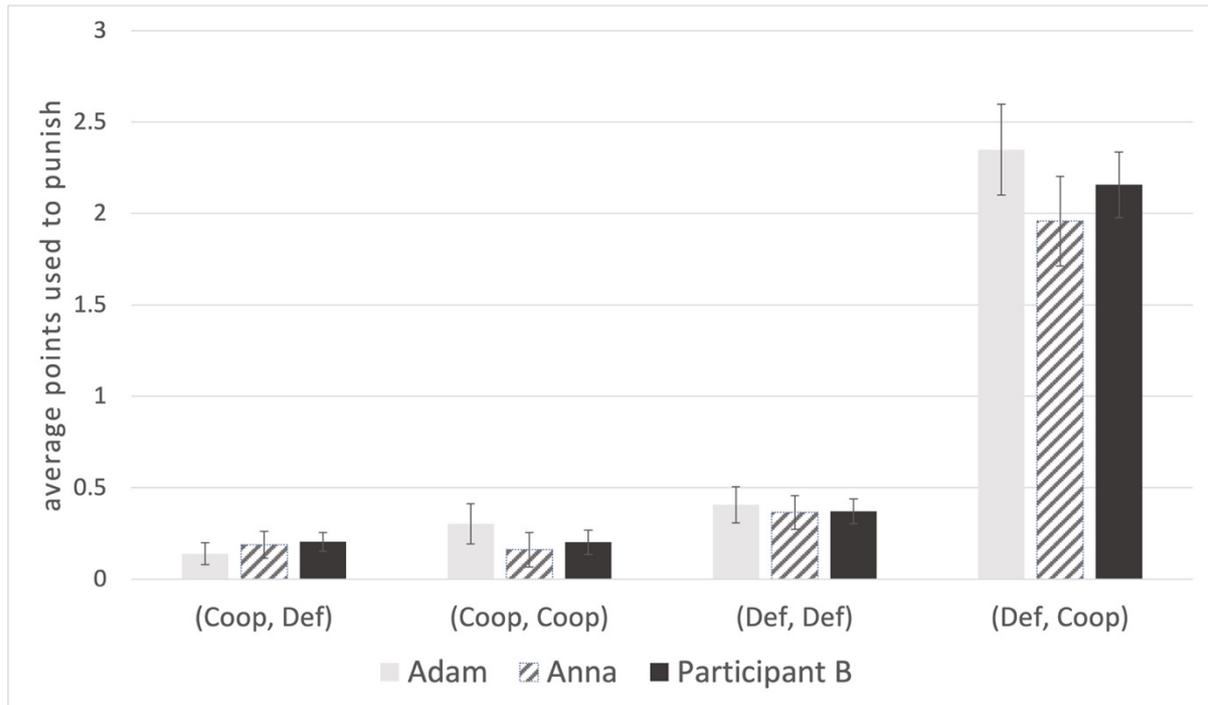

*Figure 1. Average number of points used by Participant C to punish Adam/Anna (light grey/patterned bars), and Participant B (dark grey), conditional on the decisions made by Adam/Anna and Participant B in the prisoner's dilemma. On the x-axis, the first component of each pair corresponds to the decision made by the participant to be punished. For example, the first light grey column represents the average number of points used by Participant C to punish Adam when he cooperates while the other participant, that is, Participant B, defects. Error bars represent the standard errors of the mean.*

### 2.2.2. Rewarding

Figure 2 reports the number of points used by Participants C to reward Adam (Anna) and Participant B, split by the prisoner's dilemma choices made by Adam (Anna) and Participant B. In the x-axis of the figure, the first component of each pair corresponds to the contribution made by the participant to be rewarded. For example, the first light grey column represents the average number of points used to reward Adam when he cooperates while the other participant defects. We clearly note that cooperators are rewarded way more than defectors, especially when their partner defects. The average number of points used to reward a cooperator when their partner defects is 4.02, compared to the average number of points used to reward a cooperator when their partner cooperates, that is, 2.07 ($b = 1.952$, $t = 7.09$, $p < 0.001$). In turn, this is greater than the average number of points used to reward a defector, both when their partner defects (that is, the average is 0.57; $b = 1.504$, $t = 6.95$, $p<0.001$) and when their partner cooperates (the average is 0.66; $b = 1.417$, $t = 6.25$, $p<0.001$). No statistically significant difference is present between these two latter cases ($p=0.604$). We also observe that people use more points to reward (average = 1.854) than to punish (average = 0.741). The difference is statistically significant ($b = 1.112$, $t = 12.30$, $p < .001$).

Coming to gender differences, we use linear regression to compare the number of points used to reward Adam/Anna with and without control on sex, age, and education (of the rewarder), and with and without adding the interaction between condition (Reward-Anna vs Reward-Adam) and gender (of the rewarder). Regression tables are reported in the Supplementary

Material, Tables A4-A5. In short, in contrast to our pre-registered hypothesis, we find no evidence that Adam and Anna are rewarded differently and no evidence that men and women reward Adam and Anna differently.

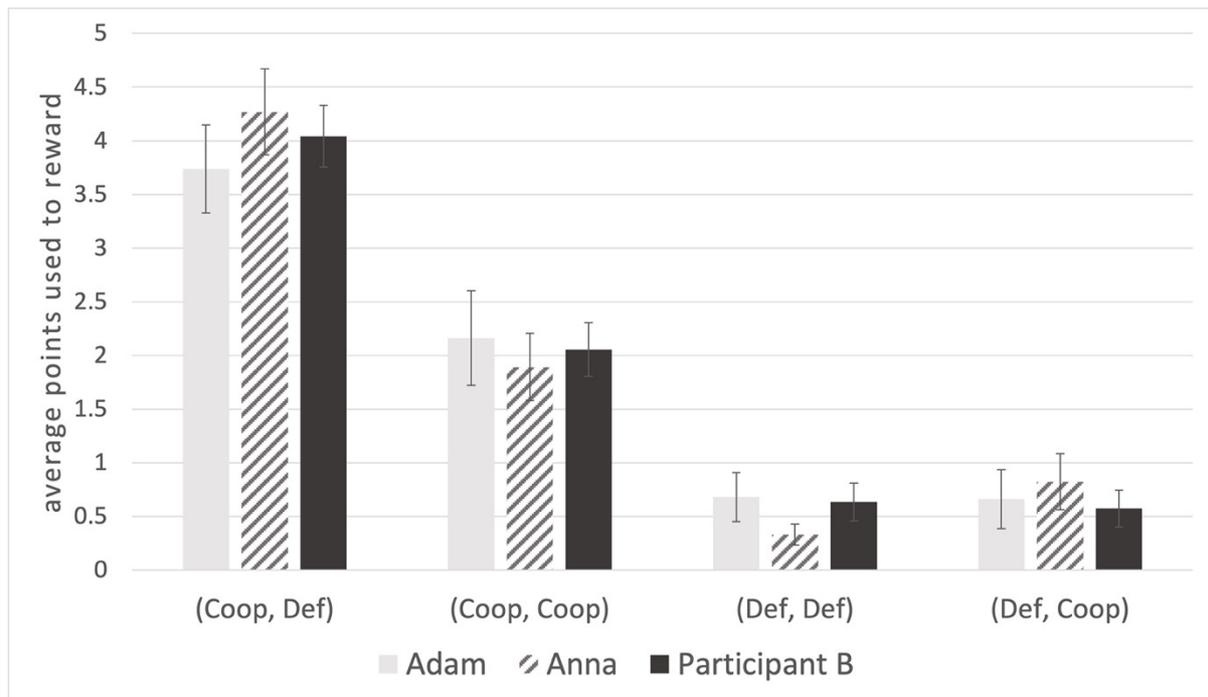

*Figure 2. Average number of points used by Participant C to reward Adam/Anna (light grey/patterned bars), and Participant B (dark grey bars), conditional on the decision made by Adam/Anna and Participant B in the prisoner's dilemma. On the x-axis, the first component of each pair corresponds to the contribution made by the participant to be rewarded. For example, the first light grey column represents the average number of points used by Participant C to reward Adam when he cooperates while the other participant, that is, Participant B, defects. Error bars represent the standard errors of the mean.*

## 3. Study 2

Study 1 provides evidence that women and men do not discriminate between genders when they have the option to punish a defector or reward a cooperator. Surprised by this result we wondered if we would find a similar result if rewarders or punishers did not have to contribute from their pool of money to reward or punish. Thus, Study 2 aims to extend this finding to a different mechanism for judging someone's behaviour. Specifically, in Study 2, participants are informed about the choice of Adam (Anna) and are asked to rate this choice on a scale of 1 to 5 stars. We believe this to be an interesting case because many online reputation systems are based on a similar rating system. This rating system was first studied in the context of social dilemmas played online by Capraro, Giardini, Vilone, and Paolucci (2016). In their paper, the authors showed that participants tend to rate cooperators higher than defectors and tend to prefer people with higher reputation as partners, even when they know nothing about how the reputation was acquired. However, they have no data regarding how people rate cooperators and defectors conditional on their gender.

### 3.1. Method

Participants were randomly divided in two conditions, the *Rate-Adam* condition and the *Rate-Anna* condition. In the *Rate-Adam* condition, participants were informed that Adam and Participant B were playing a game. The game and the instructions were identical to the ones in Study 1. After the instructions of the game, participants were asked four comprehension questions. Participants failing any comprehension question were automatically excluded from the survey. After the comprehension questions, participants were asked, in random order, to answer the following questions: (i) "Assume that Adam decided to contribute. How would you rate his behaviour? (Available answers: 1 star/2 stars/3 stars/4 stars/5 stars); and (ii) "Assume that Adam decided to not contribute. How would you rate his behaviour? (Available answers: 1 star/2 stars/3 stars/4 stars/5 stars). After rating Adam, participants entered the demographic questionnaire, after which they were given the completion code needed to claim for their payment. Analogous instructions were given in the *Rate-Anna* condition. After the survey was over, participants were paid their participation fee (30 cents). Note that the decisions in this study were not incentivised. Participants in the pre-study were not informed about the rating, therefore the rating had neither any material nor psychological effect. The method and the analysis were pre-registered at: https://aspredicted.org/blind.php?x=kr98xr.

### 3.2. Results

Whenever we found repeated IP addresses or repeated Turk IDs, we kept only the first observation, as determined by the starting date, and we discarded the rest. In doing so, we were left with a total of N=253 participants (45% females; mean age = 37.24).[10]

The average rating received by Adam when he cooperates is virtually identical to the average rating received by Anna, when she cooperates (4.22 vs. 4.21; linear regression[11]: b = 0.010, t = 0.08, p = 0.935). Similarly, the average rating received by Adam when he defects is virtually identical to the average rating received by Anna, when she defects (2.92 vs. 2.88; linear regression: b = 0.035, t = 0.24, p = 0.814). Also, splitting the sample by gender, we find that the average rating given to Adam by men when Adam cooperates is 4.19, very similar to the average rating given by women, which is 4.24 (b = 0.059, t = 0.45, p = 0.724). Similarly, the average rating given to Adam by men, when Adam defects is 2.96, close to the average rating given by women, which is 2.86 (b = -0.111, t = -0.51, p = 0.609). Finally, the average rating given to Anna by men when Anna cooperates is not statistically different from the average rating given to Anna by women (4.24 vs. 4.17; b = -0.070, t = -0.43, p = 0.669); the same happens with the average rating given to Anna when she defects (2.73 by men vs. 3.07 by women; b = 0.334, t = 1.60, p = 0.112).

### 4. Studies 3a-3b

A potential limitation of the previous studies is that the instructions contained sentences such as "Anna and Participant B…" This asymmetry might have led some participants to wonder why we gave a name to only one of the two PD participants. Some of these participants might have guessed that our experiment was about gender and this might have led them to suppress a potential gender bias. To address this issue, we conducted two variants of Study 2, where we replaced Participant B with a gender-neutral name. In Study 3a, we used Chris as a gender-neutral name; in Study 3b we used Skyler. In line with Study 1 and Study 2, these studies found that participants tend to rate Adam and Anna similarly, at least when they have no specific beliefs about the gender of Chris/Skyler. We refer to the Supplementary Material, Part 1, for the details.

---
[10] The original sample size was N = 646. A total of 61% of the observations were eliminated due to lack of comprehension or repeated Turk IDs or IP addresses.
[11] All the results presented in this section are robust to using rank-sum test.

## 5. Discussion

Across four experiments, we found evidence that people do not discriminate between men and women when they have the option to punish a defector or reward a cooperator, at least in our idealised setting of an economic experiment where cooperative behaviour is measured through a one-shot prisoner's dilemma and at least when people do not have specific beliefs about the gender of the partner of the defector/cooperator. This held not only when punishment and reward were implemented using standard laboratory techniques (decrease/increase the payoff of the defector/cooperator; Study 1), but also when they were implemented using a technique inspired by the recent development of online rating systems (rating the behaviour of a defector/cooperator from 1 to 5 stars; Study 2, Study 3a, Study 3b). The results also held across genders: neither men nor women discriminated between genders when punishing/rewarding defectors/cooperators.

This work goes beyond previous literature along several dimensions. As discussed in the Introduction, previous research has mainly focused on whether men and women punish or reward differently. While this is an important and interesting case, it does not cover the mirror question of whether men and women are punished/rewarded differently. This latter question is also interesting, in particular because of the current discussion on gender parity. We do not hide, in fact, that this is what motivated us to start this research project. To date, many experiments have shown that women are penalized more than men when they display certain behaviours in the workplace; these behaviours include self-promotion, unhelping, and initiation of negotiations; female instructors also receive lower ratings than their male counterparts. Although these are all very important situations, they do not cover a case that we believe to be very important, not only in the workplace, but in every human society: cooperation. Are women punished more than men when they fail to cooperate? Are men rewarded more than women when they cooperate? In this paper, we moved a step towards addressing these questions. Of course, these questions can be asked in general, without any reference to the specific setting of the workplace. For the sake of generality, therefore, we placed ourselves in the idealised setting of an economic experiment; we did so to avoid any particular frame that might cause a loss of generality. In this general aseptic setting, our results speak quite clearly: people do not discriminate between genders when they have the option of punishing (rewarding) defectors (cooperators), at least when they do not have beliefs about the gender of the partner of the participant to be punished/rewarded.

But, of course, generality comes with a cost. And the cost is precisely the fact that perhaps we found no gender discrimination exactly because we placed the participants in the idealised setting of an online experiment. Had the participants been placed in a prototypically feminine cooperative setting, perhaps we would have observed women being punished more than men; symmetrically, had the participants been placed in a prototypically masculine cooperative context, perhaps, this time, we would have observed men being punished more than women. Consistent with this view, a neuroeconomic article found greater neural reactions when women (vs. men) violate feminine-coded norms, and when men (vs. women) violate masculine-coded norms (Chawla, Earp, & Crockett, in press). This suggests a direction for future research that we believe to be of primary interest, that is, manipulating the frame of the prisoner's dilemma in order to make it more perceivably feminine vs. masculine, and study whether there is an asymmetry in the way the two genders are punished in "their own" prototypical prisoner's dilemma. This, in turn, opens the sub-question of how one can frame the games in such a way to make them more gender salient. In one of her many illuminating articles, Alice Eagly (2009) suggests that pro-sociality among women is primarily directed towards other people, while pro-sociality among men is primarily directed towards groups and organisations. Therefore, perhaps one can develop prisoner's dilemmas with special frames that manipulate the saliency of who receives the benefit of the cooperative action, other people vs. the group. This is certainly worth exploring in future work.

Another potential limitation of our work regards the way we manipulated the gender of the participant to be punished/rewarded, that is, by having punishers/rewarders reading the instructions of the prisoner's dilemma where the decision makers were called Adam (Anna) and Participant B. This methodology can be criticised along (at least) two dimensions. First, it is possible that some participants did not know that Anna represents a feminine name while Adam is a masculine name. Second, it is possible that some participants did not pay enough attention, and perhaps did not even read the names of the decision makers. Although we acknowledge these limitations, we do believe they are minor for two reasons. First, we carefully wrote the instructions of the experiments with the goal of referring to the gender of the decision maker, either through the direct name, or through the pronouns, in several occasions; including in the comprehension questions. For example, one of the comprehension questions of Study 1 was "In Stage 1, what is the choice that Anna should make in order to maximize her number of points?" This question contains both the name "Anna" and the pronoun "her". Taking into account that only participants who passed *all* the comprehension questions were allowed to complete the experiment, it is hard to believe that these participants did not understand the gender of the person to be punished/rewarded. Second, in Studies 3a-3b participants were asked what they believe the gender of Anna/Adam were: over 90% of the participants had correct beliefs; moreover, the main results remained qualitatively the same when restricting the analysis to these participants.

One more limitation of our results comes from the choice of recruiting subjects on AMT. This choice does have some upsides - samples collected using AMT are more heterogenous than the student sample of the classical laboratory experiment (Berinski, Huber, & Lenz, 2012) – but it does also have some downsides: subjects participate in the experiment online from their home, with little control from the experimenter; this has the consequence that some participants pay little attention to the experiment and, even worse, some try to participate multiple times in the same experiment. To avoid these problems, we followed all the standard practices to increase data quality: we asked several comprehension questions and, in situations where we found repeated IP addresses or repeated Turk IDs, we kept only the first observation and discarded the rest (Horton, Rand, & Zeckhauser, 2011; Berinsky, Margolis, & Sances, 2014; Litman & Robinson, 2020). Another potential limitation of our study is the use of small stakes. We believe this to be a minor issue. Previous work found very little stake effect in similar economic games, such as the dictator game and the trust game (Forsythe et al. 1994; Carpenter et al. 2005; Johansson-Stenman et al. 2006), at least for stakes that are not of the order of magnitude of a month of salary (Andersen et al. 2011); moreover, the average cooperation in the baseline of our pre-study was 54%, very similar to the average cooperation observed in standard laboratory experiments with larger stakes (Camerer, 2011).

A final limitation of our study is that it focuses only on third-party punishment (reward). Although this is a very well-studied (Fehr & Fischbacher, 2004; Jordan et al. 2016) and realistic form of punishment (reward), it is certainly not the only one. Another important form of punishment (reward) is peer punishment (reward). Our results are silent regarding whether peers punish (reward) women and men to a different extent when they defect (cooperate). The recent work of Gürerk et al. (2018) suggests that there might not be significant gender discrimination also in this case. However, we note that there is an important methodological difference between our work and that of Gürerk et al.: our game is one-shot, while theirs is repeated, although with potentially different groups.

In sum, our results show that people do not discriminate between genders when they have the option to punish (reward) defectors (cooperators) in a one-shot prisoner's dilemma and when they do not know the gender of the other Participant. Future work should explore what happens in gender-framed cooperation games.

## Acknowledgments

The authors thank the Editor and two anonymous referees for their detailed comments and suggestions that contributed to materially improve the content of this paper.

## References

Andersen, S., Ertaç, S., Gneezy, U., Hoffman, M., & List, J. A. (2011). Stakes matter in ultimatum games. *American Economic Review*, *101*, 3427-39.
Andreoni, J., Harbaugh, W., & Vesterlund, L. (2003). The carrot or the stick: Rewards, punishments, and cooperation. *American Economic Review*, *93*, 893-902.
Babcock, L., Recalde, M. P., Vesterlund, L., & Weingart, L. (2017). Gender differences in accepting and receiving requests for tasks with low promotability. *American Economic Review*, *107*, 714-47.
Balliet, D., Li, N. P., Macfarlan, S. J., & Van Vugt, M. (2011). Sex differences in cooperation: a meta-analytic review of social dilemmas. *Psychological Bulletin*, *137*, 881-909.
Balliet, D., Mulder, L. B., & Van Lange, P. A. (2011). Reward, punishment, and cooperation: A meta-analysis. *Psychological Bulletin*, *137*, 594-615.
Berinsky, A. J., Huber, G. A., & Lenz, G. S. (2012). Evaluating online labor markets for experimental research: Amazon. com's Mechanical Turk. *Political Analysis*, *20*, 351-368.
Berinsky, A. J., Margolis, M. F., & Sances, M. W. (2014). Separating the shirkers from the workers? Making sure respondents pay attention on self-administered surveys. *American Journal of Political Science*, *58*, 739-753.
Boring, A. (2017). Gender biases in student evaluations of teaching. *Journal of Public Economics*, *145*, 27-41.
Bowles, H. R., Babcock, L., & Lai, L. (2007). Social incentives for gender differences in the propensity to initiate negotiations: Sometimes it does hurt to ask. *Organizational Behavior and Human Decision Processes*, *103*, 84-103.
Boyd, R., & Richerson, P. J. (1992). Punishment allows the evolution of cooperation (or anything else) in sizable groups. *Ethology and Sociobiology*, *13*, 171-195.
Bueno-Guerra, N., Leiva, D., Colell, M., & Call, J. (2016). Do sex and age affect strategic behavior and inequity aversion in children? *Journal of Experimental Child Psychology*, *150*, 285-300.
Burnham, T. C. (2018). Gender, Punishment, and Cooperation: Men hurt others to advance their interests. *Socius*, *4*, 2378023117742245.
Camerer, C. F. (2011). *Behavioral game theory: Experiments in strategic interaction*. Princeton University Press.
Capraro, V., Giardini, F., Vilone, D., & Paolucci, M. (2016). Partner selection supported by opaque reputation promotes cooperative behavior. *Judgment and Decision Making*, *11*, 589-600.
Capraro, V., Jordan, J. J., & Rand, D. G. (2014). Heuristics guide the implementation of social

preferences in one-shot Prisoner's Dilemma experiments. *Scientific Reports*, *4*, 6790.

Carpenter, J., Verhoogen, E., & Burks, S. (2005). The effect of stakes in distribution experiments. *Economics Letters*, *86*, 393-398.

Chawla, M., Earp, B. D., & Crockett, M. J. (2020). A neuroeconomic framework for investigating gender disparities in moralistic punishment. *Current Opinion in Behavioral Sciences*, *34*, 166-172.

Crockett, M. J., Clark, L., Lieberman, M. D., Tabibnia, G., & Robbins, T. W. (2010). Impulsive choice and altruistic punishment are correlated and increase in tandem with serotonin depletion. *Emotion*, *10*, 855-862.

Eagly, A. H. (2009). The his and hers of prosocial behavior: An examination of the social psychology of gender. *American Psychologist*, *64*, 644-658.

Eckel, C. C., & Grossman, P. J. (2001). Chivalry and solidarity in ultimatum games. *Economic Inquiry*, *39*, 171-188.

Fehr, E., & Fischbacher, U. (2004). Third-party punishment and social norms. *Evolution and Human Behavior*, *25*, 63-87.

Fehr, E., & Gächter, S. (2002). Altruistic punishment in humans. *Nature*, *415*, 137-140.

Forsythe, R., Horowitz, J. L., Savin, N. E., & Sefton, M. (1994). Fairness in simple bargaining experiments. *Games and Economic Behavior*, *6*, 347-369.

García-Gallego, A., Georgantzís, N., & Jaramillo-Gutiérrez, A. (2012). Gender differences in ultimatum games: Despite rather than due to risk attitudes. *Journal of Economic Behavior & Organization*, *83*, 42-49.

Gürerk, Ö., Irlenbusch, B., & Rockenbach, B. (2006). The competitive advantage of sanctioning institutions. *Science*, *312*, 108-111.

Gürerk, Ö., Irlenbusch, B., & Rockenbach, B. (2018). Endogenously Emerging Gender Pay Gap in an Experimental Teamwork Setting. *Games*, *9*, 98.

Heilman, M. E., & Chen, J. J. (2005). Same behavior, different consequences: reactions to men's and women's altruistic citizenship behavior. *Journal of Applied Psychology*, *90*, 431-441.

Herrmann, B., Thöni, C., & Gächter, S. (2008). Antisocial punishment across societies. *Science*, *319*, 1362-1367.

Hilbe, C., & Sigmund, K. (2010). Incentives and opportunism: from the carrot to the stick. *Proceedings of the Royal Society B: Biological Sciences*, *277*, 2427-2433.

Johansson-Stenman, O., Mahmud, M., & Martinsson, P. (2005). Does stake size matter in trust games? *Economics letters*, *88*, 365-369.

Jordan, J. J., Hoffman, M., Bloom, P., & Rand, D. G. (2016). Third-party punishment as a costly signal of trustworthiness. *Nature*, *530*, 473-476.

Horton, J. J., Rand, D. G., & Zeckhauser, R. J. (2011). The online laboratory: Conducting experiments in a real labor market. *Experimental Economics*, *14*, 399-425.

Litman, L., & Robinson, J. (2020). *Conducting Online Research on Amazon Mechanical Turk and Beyond*. SAGE Publications, Incorporated.

MacNell, L., Driscoll, A., & Hunt, A. N. (2015). What's in a name: Exposing gender bias in student ratings of teaching. *Innovative Higher Education*, *40*, 291-303.

Mengel, F., Sauermann, J., & Zölitz, U. (2019). Gender bias in teaching evaluations. *Journal of the European Economic Association*, *17*, 535-566.

Milinski, M., Semmann, D., Krambeck, H. J., & Marotzke, J. (2006). Stabilizing the Earth's climate is not a losing game: Supporting evidence from public goods experiments. *Proceedings of the National Academy of Sciences*, *103*, 3994-3998.

Nowak, M. A. (2006). Five rules for the evolution of cooperation. *Science*, *314*, 1560-1563.

Paolacci, G., Chandler, J., & Ipeirotis, P. G. (2010). Running experiments on amazon mechanical turk. *Judgment and Decision making*, *5*, 411-419.

Perc, M., Jordan, J. J., Rand, D. G., Wang, Z., Boccaletti, S., & Szolnoki, A. (2017). Statistical physics of human cooperation. *Physics Reports*, *687*, 1-51.

Peysakhovich, A., Nowak, M. A., & Rand, D. G. (2014). Humans display a 'cooperative phenotype' that is domain general and temporally stable. *Nature Communications*, *5*, 4939.

Rand, D. G. (2017). Social dilemma cooperation (unlike Dictator Game giving) is intuitive for men as well as women. *Journal of Experimental Social Psychology*, *73*, 164-168.

Rudman, L. A. (1998). Self-promotion as a risk factor for women: the costs and benefits of countersterotypical impression management. *Journal of Personality and Social Psychology*, *74*, 629-645.

Solnick, S. J. (2001). Gender differences in the ultimatum game. *Economic Inquiry*, *39*, 189-200.

# Supplementary Material to the paper "Punishing defectors and rewarding cooperators: Do people discriminate between genders?"

Valerio Capraro and Hélène Barcelo

## Abstract

This Supplementary Material is divided in five parts. Part 1 details the results of the Pre-study, Study 3a, and Study 3b. Part 2 describes the computation of participants' bonuses in the Pre-study and in Study 1. Part 3 reports some supplementary analyses, including the regression tables of Study 1 and Study 2. Part 4 presents the analysis of the beliefs. Part 5 shows the experimental instructions.

# Part 1. Supplementary studies

**Pre-study**

In this preliminary study, we collect data regarding how participants play a prisoner's dilemma (PD) knowing that they might be rewarded or punished by a third party. This study serves two goals. First, it allows us to avoid deception in Study 1, where participants will be asked to punish/reward the behaviour of participants who played the PD in the pre-study. Therefore, in order to avoid deception, we need participants who actually play the PD knowing that they are observed and might be punished or rewarded later on. The second goal of the pre-study is to compare the efficacy of punishment with the efficacy of reward in promoting cooperation in a one-shot prisoner's dilemma. This is interesting in itself, because it allows us to contribute to the literature comparing the effectiveness of these two mechanisms for promoting cooperation (Balliet, Mulder, & Van Lange, 2011). In the pre-study, we also collect data about participants' beliefs about men's and women's cooperation levels. We do so because hypothesis H2 (our pre-registered hypothesis) is based on the underlying assumption that women are expected to cooperate more than men. This part of the study is reported in the Supplementary Material, Part 3.

*Method*

The experiment was conducted on AMT. Participants were randomly assigned to one of three conditions. In the *PD-Baseline* condition, they made a decision in a standard two-player, one-shot, prisoner's dilemma (see below for the details); in the *PD-Punishment* condition, they made a decision in the same prisoner's dilemma as in the baseline condition, but knowing that a third party will be informed about their decision and will have the opportunity to punish them; in the *PD-Reward* condition, they made a decision in the same prisoner's dilemma as in the baseline condition, but knowing that a third party will be informed about their decision and will have the opportunity to reward them. Specifically, in the *PD-Baseline* condition each participant was informed that he or she would be paired with another participant, who was reading the same set of instructions. The two participants were named Participant A and Participant B. (All participants of the pre-study were told that they were Participant A in the game; we did this to make the instructions easier to understand and, since the game is symmetric, it has no impact on the study.) Participants were informed that they started with 20 points. They were also informed that there was a common pool, with 0 points at the beginning of the experiment. The participants were told that they could choose to contribute, or not, all of their points to the common pool. No intermediate contributions were allowed.[12] The total amount in the common pool was then multiplied by 1.5 and equally distributed between the two participants. This payoff structure implies that it is individually optimal for both participants to not contribute to the common pool; however, if both participants do not contribute, then they receive a payoff of 20 points, which is less than what they would have gotten if they had both contributed (30 points). In the *PD-Punishment* condition, the participants were additionally informed of the existence of a third participant,

---

[12] If we allow intermediate contributions in the pre-study, then the complexity of Study 1 would be much (and unnecessarily) higher, because, for every possible profile of contributions, we would have to ask Participant C to punish (reward) each of the two participants (Anna/Adam and Participant B). Assuming that, as intermediate contributions we allow any integer between 0 and 20, this would lead to $21^2 = 441$ profiles of contributions, compared to only 4, as they are in our case. Therefore, to avoid this unnecessary increase in complexity, we decided to allow only extreme contributions. Restricting the contributions to two extremes implies that our two-player public good game has the structure of a prisoner's dilemma.

C, who would be told the decision of both Participant A and Participant B, and could then use their points to deduct points from either Participant A or Participant B, or both (see Study 1). The *PD-Reward* condition was identical to the *PD-Punishment* condition, with the only difference that the participants were informed that Participant C could use their points to add points to their account.

Participants were informed that 1 point would correspond to 1 US cent. After reading the instructions, participants were asked some comprehension questions. Participants failing the comprehension questions were automatically excluded from the survey and received no payment; this exclusion criterium was described at the beginning of the survey (see experimental instruction in the Supplementary Material, Part 5). Participants passing the comprehension questions were allowed to make their prisoner's dilemma decision, after which they entered a demographic questionnaire, at the end of which they received the completion code needed to claim their payment through AMT. After the experiment ended and after conducting Study 1, we computed the bonuses and we paid them in addition to the participation fee (50 cents). In this way, no deception occurred (see Supplementary Material, Part 2, for details about the computation of the bonuses). The design, the exclusion criteria, and the analysis were pre-registered at the same link as Study 1: http://aspredicted.org/blind.php?x=tc76r2.

### *Results*

As pre-registered, whenever we find more than one observation with the same IP address or Turk ID, we keep only the first observation, as determined by the starting date, and discard the rest. In doing so, we were left with a total of N=526 participants (47% females; mean age = 36.4).[13] Comparing the rate of cooperation in the baseline with the rate of cooperation in the two treatments, we clearly see that both mechanisms (punishment and reward) increased cooperation. Specifically, in the *PD-baseline* condition, the rate of cooperation was 54.3% (in line with previous works using a similar payoff structure; Capraro, Jordan, & Rand, 2014), whereas in the *PD-Punishment* condition and in the *PD-Reward* condition, the rates of cooperation were 72.9% and 71.5%, respectively; both significantly higher than the rate of cooperation in the *PD-baseline* condition (logistic regression; *PD-Punishment* vs *PD-Baseline*: b = 0.819, z = 3.30, p = 0.001; *PD-Reward* vs *PD-Baseline*: b = 0.751, z = 3.35, p = 0.001). There was no statistical difference between the rate of cooperation in the *PD-Punishment* condition and the rate of cooperation in the *PD-Reward* condition (p = 0.814). This is in line with the meta-analysis by Balliet, Mulder, and Van Lange (2011), finding that the punishment effect in one-shot cooperation games is statistically the same as the reward effect. We also note that the very clear effect of punishment and reward on contribution rates also shows that a substantial proportion of participants understood the role of Participant C.

**Study 3a**

### *Method*

---

[13] The original sample size was N=1,538 participants. Among these, about 63% participants were automatically excluded from the survey because they did not pass the comprehension questions, leaving us with a sample size of N=596; the other 70 observations were eliminated because of repeated IP addresses and repeated Turk IDs. An exclusion rate of 63% because of lack of comprehension is in line with previous research. For example, Capraro, Jordan and Rand (2014) report an exclusion rate of 40% in a prisoner's dilemma with 4 comprehension questions; in our case two out of 3 treatments, had 7 comprehension questions, to make sure that people understood the punishment/reward phase, the baseline had 4 comprehension questions.

Study 3a is similar to Study 2, albeit for two main differences. First of all, Participant B is replaced by Chris. Second, after participants make the rating decision, they are asked the following two questions in random order: "While answering the questions, what gender did you think Anna (Adam, depending on the condition) was?", "While answering the questions, what gender did you think Chris was?" The available options were, in random order: male, female, neutral, I didn't know or I didn't think about it. This study was pre-registered at: https://aspredicted.org/kx9fj.pdf.

*Results*

After eliminating repeated IP addresses or repeated Turk IDs, we were left with N=331 participants (43% females; mean age = 38.40).[14]

The average rating received by Adam when he cooperates is virtually identical to the average rating received by Anna, when she cooperates (4.46 vs. 4.49; linear regression[15]: b = -0.026, t = 0.29, p = 0.775). However, the average rating received by Adam when he defects is significantly smaller than the average rating received by Anna, when she defects (2.12 vs. 2.40; linear regression: b = -0.283, t = -2.47, p = 0.014). Splitting the sample by gender, we find that the average rating given to Adam by men when Adam cooperates is 4.47, which is very similar to the average rating given by women, which is 4.44 (b = 0.029, t = 0.21, p = 0.831). Similarly, the average rating given to Adam by men, when Adam defects is 2.11, is very similar to the average rating given by women, which is 2.11 (b = 0.003, t = 0.02, p = 0.985). Finally, the average rating given to Anna by men when Anna cooperates is similar to the average rating given to Anna by women (4.52 vs. 4.44; b = 0.087, t = 0.75, p = 0.455); the same happens with the average rating given to Anna when she defects (2.38 by men vs. 2.43 by women; b = 0.057, t = 0.36, p = 0.721).

As pre-registered, as a robustness check, we repeated the analysis controlling for a categorical variable representing what participants believe to be Chris' gender and by restricting the analysis to those subjects who believe that Anna is a female (96% of the participants) and Adam is a male (92% of the participants), with and without control on what participants believe to be Chris' gender. The previous results remain qualitatively the same.

As an exploratory analysis, we noticed that the vast majority (84.6%) of subjects believed that Chris was a male. Only 2.7% of the subjects believed that Chris was a female. If we restrict the main analysis to subjects who report that Chris was gender-neutral or report that they didn't know or didn't think about Chris' gender, we found that the average rating received by Anna when she cooperates is virtually identical to the average rating received by Adam when he cooperates (4.09 vs 4.14; b = 0.048, t = 0.14, p=0.887), and also the average rating received by Anna when she defects is virtually identical to the average rating received by Adam when he defects (2.48 vs 2.52; b = 0.048, t = 0.13, p=0.899). This exploratory analysis suggests that the fact that Adam is punished more than Anna when he defects is entirely driven by the fact that people believe that Chris is a male, and that no significant gender differences in rating are present when people have no beliefs about the gender of the other participant. The next study attempts to clarify this point.

---

[14] The original sample size was N = 534. A total of 38% of the observations were eliminated due to lack of comprehension or repeated Turk IDs or IP addresses.
[15] All the results presented in this section are robust to using rank-sum test.

**Study 3b**

The limitation of Study 3a was that, to our surprise, most of the participants believed that Chris was a male. To find a more gender-neutral name, we consulted a registry of American gender-neutral names. We found that the most used gender-neutral names are Charlie (50% females, 50% males), Finley (58% females, 42% males), and Skyler (54% females, 46% males).[16] After consulting with several Americans, we opted for Skyler.

*Method*

Study 3b is identical to Study 3a, with only one difference: the name Chris was replaced by Skyler. We also slightly changed the pre-registration, where we now report also that we would test our main hypothesis by restricting the analysis to subjects who believe that Skyler is gender-neutral or that they didn't know or didn't think about Skyler's gender. This study was pre-registered at: https://aspredicted.org/kd4t3.pdf.

*Results*

After eliminating repeated IP addresses or repeated Turk IDs, we were left with N=310 participants (46% females; mean age = 39.78).[17]

The average rating received by Adam when he cooperates is significantly smaller than the average rating received by Anna, when she cooperates (4.26 vs. 4.46; linear regression[18]: $b = -0.202$, $t = -2.06$, $p = 0.040$). The average rating received by Adam when he defects is similar to the average rating received by Anna, when she defects (2.57 vs. 2.46; linear regression: $b = 0.109$, $t = 0.83$, $p = 0.407$). Splitting the sample by gender, we find that the average rating given to Adam by men when Adam cooperates is 4.25, which is very similar to the average rating given by women, which is 4.30 ($b = -0.030$, $t = -0.21$, $p = 0.832$). Analogously, the average rating given to Adam by men, when Adam defects is 2.53, is very similar to the average rating given by women, which is 2.62 ($b = 0.100$, $t = 0.59$, $p = 0.559$).
Finally, the average rating given to Anna by men when Anna cooperates is not statistically different from the average rating given to Anna by women (4.39 vs. 4.52; $b = 0.133$, $t = 1.03$, $p = 0.306$); the same happens with the average rating given to Anna when she defects (2.46 by men vs. 2.46 by women; $b = -0.001$, $t = -0.01$, $p = 0.995$).

As pre-registered, as a robustness check, we repeated the analysis controlling for a categorical variable representing what participants believe to be Skyler's gender and by restricting the analysis to those subjects who believe that Anna is a female (97% of the participants) and Adam is a male (96% of the participants), with and without control on what participants believe to be Skyler's gender (31% of the participants believed that Skyler was a man, 47% of the participants believed that Skyler was a woman, 22% of the participants thought that Skyler was gender-neutral or they didn't know or didn't think about Skyler's gender). The previous results remain qualitatively the same.

Finally, as pre-registered, we repeat the main analysis by restricting it to those participants who report that Skyler is gender-neutral or that they didn't know or didn't think about

---

[16] https://www.mother.ly/child/top-50-gender-neutral-baby-names-youll-obsess-over-
[17] The original sample size was N = 552. A total of 44% of the observations were eliminated due to lack of comprehension or repeated Turk IDs or IP addresses.
[18] All the results presented in this section are robust to using rank-sum test.

Skyler's gender. In line with Study 3a, we found that, among these participants, the average rating received by Adam when he cooperates is similar to the average rating received by Anna, when she cooperates (4.33 vs. 4.62; b = -0.299, t = -1.56, p = 0.125). (Note, however, that this result should be taken with caution, because the difference 4.62 – 4.33 is actually higher than the significant difference in the whole sample, 4.46 – 4.26; therefore, it is possible that the lack of significance is due to a small sample size.) Similarly, the average rating received by Adam when he defects is similar to the average rating received by Anna, when she defects (2.56 vs. 2.62; b = -0.067, t = -0.25, p = 0.804).

As a non-preregistered, exploratory analysis, we also put together the results of Study 3a and Study 3b, to confirm that subjects who believe that Chris or Skyler is gender neutral or didn't think about it, do not discriminate among genders when they are asked to rate Anna or Adam. Indeed, among these subjects, the average rating given to Adam when he cooperates is similar to the average rating given to Anna when she cooperates (4.26 vs 4.38, b = -0.112, t = -0.64, p = 0.524). Similarly, among these subjects, the average rating given to Adam when he defects is similar to the average rating given to Anna when she defects (2.55 vs 2.56, b = -0.009, t = -0.04, p = 0.968).[19]

In sum, both Study 3a and Study 3b show that, when participants have no specific beliefs about the gender of Adam (Anna)'s partner, they tend to rate Adam and Anna similarly, both when they defect and when they cooperate. This is in line with Study 1 and Study 2.

Regarding the effect of the gender of the partner of the participant to be punished/rewarded. Study 3a found that participants who believe that Chris is a male tend to rate Adam for defection lower than they rate Anna. However, this result was not replicated in Study 3b (but the trend is in the same direction; see Footnote 19). On the other hand, Study 3b found that participants who believe that Skyler is a female tend to rate Adam, when he defects, marginally significantly higher than Anna (see Footnote 19). Together, these trends seem to suggest that men might be rated more harshly than women when they defect against a man, but women tend to be rated more harshly than men when they defect against a woman. This, however, should be taken with caution, because of the inconsistencies across studies and because the multiple splits of the sample raise the issue of multiple hypothesis testing. Further work should explore this issue in greater depth.

### Part 2. Computation of the bonuses of the pre-study and Study 1

The computation of the bonuses of the pre-study and Study 1 was not straightforward. In order to guarantee that each decision was incentivised, we proceeded as follows. To facilitate the description of the procedure we explain it for the Punish case. The reward case was similar. Each participant of the pre-study *Punish* condition was first paired with another participant of the pre-study *Punish* condition, therefore creating a pool of Participant A and

---

[19] As additional exploratory analysis, we also looked at the ratings received by Adam vs Anna, as a function of what people believe to be Skyler's gender. When we restrict the analysis to participants who believe that Skyler is a male (N=97), we find that these participants rate Adam, when he defects, on average 2.41, compared to the 2.61 assigned to Anna (b = -0.197, t = -0.83, p = 0.409). The same participants rate Adam, when he cooperates, on average 4.27, compared to 4.46 assigned to Anna (b = -0.182, t = -1.13, p = 0.261). If we restrict the analysis to participants who believe that Skyler is a female (N=146), we find that these participants rate Adam, when he defects, on average 2.70, compared to the 2.33 assigned to Anna (b = 0.372, t = 1.91, p = 0.059). The same participants rate Adam, when he cooperates, on average 4.21, compared to 4.42 assigned to Anna (b = -0.209, t = -1.31, p = 0.192).

Participant B having played the game, henceforth (A,B). If the number of participants of the pre-study punish condition was *2n*, then *n* pairs were created; if the number was *2n+1*, then the last participant (alphabetically, using the TurkID) was paired with the first participant, for a total of *n+1* pairs. (In this case, the payoff of the first participant was computed by selecting at random the decision of the second participant or that of the last participant.)

Next, the paired participants (A,B) were randomly matched with a participant of the Study 1 *Punish-Anna* or *Adam* conditions depending on the gender of A. That is, if Participant A of the pre-study was a woman, that pair (A,B) was matched with a participant C of the Study 1 *Punish-Anna* condition. If Participant A was a male, then that pair (A,B) was matched with a Participant C of the Study 1 *Punish-Adam* condition.

We kept this procedure going until we ran through all participants of Study 1. Since Study 1 had more participants than there were pairs in the pre-study, some pairs of participants of the pre-study were matched with more than one participant in Study 1. Therefore, to compute the bonuses of the participants in the pre-study, we took the average punishment implemented by the participants with whom they were matched. In situations when this operation gave rise to a decimal punishment (or reward), we used the floor function. In this way, we assure that all decisions (in both the pre-study and Study 1) were incentive compatible.

## Part 3. Supplementary analysis

**Table A1**

| Study | Condition | N | Female |
|---|---|---|---|
| Pre-study | Baseline | 282 | 0.46 |
| | Punishment | 107 | 0.46 |
| | Reward | 137 | 0.50 |
| Study 1 | Punish-Adam | 157 | 0.45 |
| | Punish-Anna | 142 | 0.52 |
| | Reward-Adam | 120 | 0.42 |
| | Reward-Anna | 125 | 0.57 |
| Study 2 | Rate-Adam | 119 | 0.47 |
| | Rate-Anna | 134 | 0.44 |
| Study 3a | Rate-Adam | 171 | 0.43 |
| | Rate-Anna | 160 | 0.43 |
| Study 3b | Rate-Adam | 161 | 0.39 |
| | Rate-Anna | 149 | 0.54 |

**Table A1** Breakdown of participants by proportion of females and condition across all studies.

**Table A2**

| | Points used to punish | | | |
|---|---|---|---|---|
| Adam | 0.26 | 0.20 | 0.13 | 0.05 |
| | (0.28) | (0.29) | (0.39) | (0.39) |
| Female punisher | | 0.09 | 0.03 | 0.00 |
| | | (0.25) | (0.29) | (0.29) |
| Adam*Female | | | 0.24 | 0.31 |
| | | | (0.57) | (0.57) |
| Age | | 0.00 | | 0.00 |
| | | (0.01) | | (0.01) |
| Education | | -0.16 | | -0.17 |
| | | (0.11) | | (0.11) |
| Constant | 2.09*** | 2.80*** | 2.06*** | 2.86*** |
| | (0.15) | (0.68) | (0.21) | (0.69) |
| R-squared | 0.00 | 0.01 | 0.00 | 0.01 |
| N | 591 | 583 | 583 | 583 |

**Table A2** Linear regressions predicting the number of points used to deduct points from Adam/Anna's account, when Adam/Anna defects and Participant B cooperates. Standard error in parentheses. Significance levels: *: $p < 0.1$, **: $p < 0.05$, ***: $p < 0.01$.

**Table A3**

| | Points used to punish | | | |
|---|---|---|---|---|
| Adam | 0.04 | 0.03 | 0.06 | 0.02 |
| | (0.11) | (0.11) | (0.15) | (0.15) |
| Female punisher | | 0.06 | 0.05 | 0.05 |
| | | (0.10) | (0.11) | (0.11) |
| Adam*Female | | | -0.00 | 0.02 |
| | | | (0.22) | (0.22) |
| Age | | -0.00 | | -0.00 |
| | | (0.00) | | (0.00) |
| Education | | -0.07* | | -0.07* |
| | | (0.04) | | (0.04) |
| Constant | 0.37*** | 0.84*** | 0.33*** | 0.85*** |
| | (0.06) | (0.26) | (0.08) | (0.27) |
| R-squared | 0.00 | 0.01 | 0.00 | 0.01 |
| N | 581 | 574 | 574 | 574 |

**Table A3** Linear regressions predicting the number of points used to deduct points from Adam/Anna's account, when Adam/Anna and Participant B both defect. Standard error in parentheses. Significance levels: *: $p < 0.1$, **: $p < 0.05$, ***: $p < 0.01$.

**Table A4**

|  | Points used to reward | | | |
|---|---|---|---|---|
| Adam | 0.29 | 0.29 | 0.20 | 0.25 |
|  | (0.43) | (0.44) | (0.58) | (0.59) |
| Female punisher |  | -0.26 | -0.33 | -0.28 |
|  |  | (0.38) | (0.44) | (0.44) |
| Adam*Female |  |  | 0.18 | 0.09 |
|  |  |  | (0.88) | (0.88) |
| Age |  | 0.02 |  | 0.02 |
|  |  | (0.02) |  | (0.02) |
| Education |  | 0.16 |  | 0.15 |
|  |  | (0.17) |  | (0.19) |
| Constant | 2.00*** | 0.54 | 2.18*** | 0.56 |
|  | (0.21) | (0.97) | (0.31) | (0.99) |
| R-squared | 0.00 | 0.01 | 0.00 | 0.01 |
| N | 478 | 470 | 470 | 470 |

**Table A4** Linear regressions predicting the number of points used to add points to Adam/Anna's account, when Adam/Anna and Participant B both cooperate. Standard error in parentheses. Significance levels: *: $p < 0.1$, **: $p < 0.05$, ***: $p < 0.01$.

**Table A5**

|  | Points used to reward | | | |
|---|---|---|---|---|
| Adam | -0.38 | -0.30 | -0.10 | -0.00 |
|  | (0.47) | (0.48) | (0.64) | (0.64) |
| Female punisher |  | -0.20 | -0.09 | -0.04 |
|  |  | (0.41) | (0.48) | (0.48) |
| Adam*Female |  |  | -0.56 | -0.67 |
|  |  |  | (0.96) | (0.96) |
| Age |  | 0.04** |  | 0.04** |
|  |  | (0.02) |  | (0.02) |
| Education |  | 0.08 |  | 0.09 |
|  |  | (0.18) |  | (0.18) |
| Constant | 4.12*** | 2.22** | 4.13*** | 1.09** |
|  | (0.23) | (1.06) | (0.34) | (1.08) |
| R-squared | 0.00 | 0.00 | 0.00 | 0.01 |
| N | 477 | 469 | 469 | 469 |

**Table A5** Linear regressions predicting the number of points used to add points to Adam/Anna's account, when Adam/Anna cooperates and Participant B defects. Standard error in parentheses. Significance levels: *: $p < 0.1$, **: $p < 0.05$, ***: $p < 0.01$.

## Part 4. Analysis of beliefs.

As mentioned in the Methods section of the pre-study, this experiment contained two more conditions, devoted to testing the pre-registered hypothesis that women are expected to cooperate more than men. In the *Guess-Adam (Guess-Anna)* condition, participants read the instructions of the prisoner's dilemma, where two participants are named *Adam (Anna)* and *Participant B*. Participants passing the comprehension questions, analogue to the comprehension questions in the baseline of the pre-study, were then are asked to guess Adam's (Anna's) contribution. Correct guesses were incentivised with a $0.40 prize. N=127

participants participated in the *Guess-Adam* condition and N=124 participants participated in the *Guess-Anna* condition. 46% of the participants in the Guess-Adam condition believed that Adam would cooperate; 56% of the participants in the Guess-Anna condition believed that Anna would cooperate. Therefore, the results were in the expected direction. However, the difference was not statistically significant, with (p=0.158) and without (p=0.144) control on sex, age, and education. However, adding the interaction between condition and gender of the decision maker reveals a significant negative interaction *Guess-Anna\*Female* and a significant positive main effect of *Guess-Anna* and *Female*. Additional analysis indeed suggests that, in contrast to our pre-registered hypothesis, we do not find clear evidence that women are expected to be more cooperative than men in general; we do find indeed that men are expected to be less cooperative than women, but only by other men. Results are summarised in Table A6 and Table A7.

**Table A6**

|  | Cooperation | | | |
| --- | --- | --- | --- | --- |
| Guess-Anna | 0.09 | 0.09 | 0.22** | 0.24*** |
|  | (0.06) | (0.46) | (0.09) | (0.09) |
| Female |  | 0.11* | 0.27*** | 0.27*** |
|  |  | (0.06) | (0.48) | (0.48) |
| Guess-Anna*Female |  |  | -0.29** | -0.31** |
|  |  |  | (0.13) | (0.13) |
| Age |  | 0.00 |  | 0.00 |
|  |  | (0.00) |  | (0.00) |
| Education |  | 0.01 |  | 0.02 |
|  |  | (0.03) |  | (0.03) |
| Constant | 0.47*** | 0.36*** | 0.35*** | 0.27** |
|  | (0.04) | (0.12) | (0.06) | (0.13) |
| R-squared | 0.01 | 0.03 | 0.04 | 0.06 |
| N | 249 | 248 | 248 | 248 |

**Table A6** Logistic regressions predicting participants' beliefs about Adam's and Anna's cooperation level (1=cooperate, 0=defect). Standard error in parentheses. Significance levels: *: p < 0.1, **: p < 0.05, ***: p < 0.01.

**Table A7**

|  | Men | Women |
| --- | --- | --- |
| Guess-Adam | 0.35 | 0.61 |
|  | (0.06) | (0.07) |
| Guess-Anna | 0.57 | 0.55 |
|  | (0.06) | (0.06) |

**Table A7** Mean expectations of Adam vs. Anna's cooperation levels, broken down by gender of the participant. Standard error in brackets.

# Part 5. Experimental instructions

**Pre-study**

We report the instructions only of the *PD-Punishment* condition. The experimental instructions of the other conditions are very similar.

*Instructions*

Welcome to this HIT.

This HIT will take you about ten minutes. For your participation in this HIT, you will earn 50 cents. You can also earn additional money depending on the decisions that you and the other participants will make.

During this HIT, we use points. We do not deal with dollars. Each income will be temporarily calculated in points. The total amount of points accumulated during the HIT will be converted to dollars at the end of the HIT. The rate of exchange is: **1 point = 1 cent.**

At the end of the HIT, you will receive the equivalent of the points earned during the HIT plus the 50 cents for participating.

This HIT has three participants: Participant A, B, and C. **You are Participant A**. There are two stages. At the beginning of Stage 1, you will be paired with Participant B, and each of you will be given a number of points. You will be asked whether you want to contribute your points to a common pool or if you want to keep the points for yourself. You and Participant B are linked with an observer, Participant C, who will learn the choices that you and Participant B will have made. In Stage 2, participant C will be able to deduct points from your and/or Participant B's accounts, depending on the choices you made in Stage 1. In the event that Participant C decides to deduct points, your account and/or that of Participant B will be reduced accordingly.

Specific instructions follow in the next screen.

IMPORTANT: to make sure you understood the decision problems, you will be asked some simple questions, each of which has only one correct answer. If you fail to correctly answer any of those questions, the survey will automatically end and you will not receive any redemption code and consequently you will not get any payment.

With this in mind, do you wish to continue? (Available answers: Continue/End)

*Page break*

Please, read carefully the following instructions.

FIRST STAGE

You (Participant A) and Participant B each start this stage with 20 points. There is also a common pool with 0 points at the beginning of the survey. You can choose to contribute, or not, all of your points to the common pool. No intermediate contributions are allowed. If you contribute 20 points to the common pool they are multiplied by 1.5; that is, the amount of points in the common pool will become 20 x 1.5 = 30 points. Participant B is reading the same set of instructions. Thus, Participant B will also be asked whether they want to contribute their 20 points to the common pool and Participant B's contribution will also be multiplied by 1.5. After you and participant B have decided whether or not to contribute to the common pool, the total amount of points in the common pool will be split equally between the two of you, whether you contributed to it or not.

Importantly, you and Participant B cannot communicate. You have to make your choice independently of each other and you cannot decide on a joint strategy.

Thus, in summary:

If both you and Participant B contribute your 20 points to the common pool, then the total amount of points in the common pool is 30 + 30 = 60 points, which must then be split equally between you and Participant B. Thus, at the end of the first stage, you and Participant B, will each have 30 points = 20 (your starting amount) - 20 (your contribution) + 30 (half of the common pool).

If you contribute your 20 points to the common pool, but Participant B does not do so, then the total amount of points in the common pool will be 30 points (0 + 30 + 0). Thus, at the end of the first stage, you will have 15 points (20 - 20 + 15), while Participant B will have 35 points (20 - 0 + 15).

Similarly, if you do not contribute your 20 points to the common pool, but participant B does, then at the end of the first stage, you will have 35 points while Participant B will have 15 points.

If neither you nor Participant B contribute your 20 points to the common pool, then, at the end of the first stage, both you and Participant B will have 20 points.

This is the end of the first stage.

SECOND STAGE

In this stage, an observer, Participant C, is introduced who will learn the decisions you and Participant B made in stage 1. The observer is given 40 points, which can be used to deduct points from your account and/or Participant B's account. For every deduction point the observer uses, their number of points will be reduced by one point while either yours or Participant B's number of points will be reduced by three.

For example, if Participant C decides to use 2 deduction points to deduct points from your account and 5 deduction points from Participant B's account, then:

Participant C will end this second stage with 40 - 2 - 5 = 33 points;

you will end it with whatever amount you got in the first stage minus 2*3 = 6 points;

and Participant B will end this stage with whatever amount they got in the first stage minus 5*3 = 15 points.

No negative amounts are allowed, so the maximum number of points that can be deducted is equal to the number of points you got in the first stage of the survey.

At the end of this stage, the survey ends. This is the only interaction between you and the other two participants. You will be paid according to the number of points you have accumulated during the two stages (plus the 50 cents for your participation in this HIT). 1 point will correspond to 1 cent.

None of the participants will ever be aware of the identities of the other participants, either during or after this survey, ensuring total anonymity of all persons involved.

Now we ask you some comprehension questions to ascertain that you understood the decision problem.

In Stage 1, what is the choice that YOU should make in order to maximize your number of points? (Available answers: Contribute the 20 points to the common pool/ Do not contribute the 20 points to the common pool)

In Stage 1, what is the choice that YOU should make in order to maximize the amount in the COMMON POOL? (Available answers: Contribute the 20 points to the common pool/ Do not contribute the 20 points to the common pool)

In Stage 1, what is the choice that Participant B should make in order to maximize their number of points? (Available answers: Contribute the 20 points to the common pool/ Do not contribute the 20 points to the common pool)

In Stage 1, what is the choice that Participant B should make in order to maximize the amount of points in the COMMON POOL? (Available answers: Contribute the 20 points to the common pool/ Do not contribute the 20 points to the common pool)

In Stage 2, what is the choice that the observer, Participant C, should make in order to maximize their own number of points? (Available answers: Do not deduct points from anyone/Deduct points from someone)

In Stage 2, what is the role of the observer, Participant C? (Available answers: Deduct points from you and/or Participant B, depending on your choices/Add points to you and/or Participant C, depending on your choices)

Assume that you end Stage 1 of the HIT with x points and that Participant B ends Stage 1 of the HIT with y points. If the observer, Participant C, decides to use 2 deduction points to deduct points from your account and decides to use 2 deduction points to deduct points from Participant B's account, how many points will you and Participant B each have at the end of Stage 2? (Available answers: You get x - 2 points, and Participant B gets y - 2 points/You get x - 6 points, and Participant B gets y - 6 points/You get x + 2 points, and Participant B gets y + 2 points/You get x + 6 points, and Participant B gets y + 6 points)

*Page break*

You passed the comprehension questions.

It is now time to make your choice. Remember that you are Participant A.

What will you do? (Available answers: Contribute the 20 points to the common pool/ Do not contribute the 20 points to the common pool)

**Study 1**

We report the instructions only of the *Punish-Anna* condition. The instructions of the other conditions are similar.

*Instructions*

Welcome to this HIT.

This HIT will take you about ten minutes. For your participation in this HIT, you will earn 50 cents. You can also earn additional money depending on the decisions that you and the other participants will make.

During this HIT, we use points. We do not deal with dollars. Each income will be temporarily calculated in points. The total amount of points accumulated during the HIT will be converted to dollars at the end of the HIT. The rate of exchange is: **1 point = 1 cent.**

At the end of the HIT, you will receive the equivalent of the points earned during the HIT plus the 50 cents for participating.

This HIT has three participants: Anna, Participant B, and Participant C. There are two stages. At the beginning of Stage 1, Anna will be paired with Participant B, and each of them will be given a number of points. They will be asked whether they want to contribute their points to a common pool or if they want to keep the points for themselves. Anna and Participant B are aware that there is an observer, Participant C, who will learn the choices that they will have made. In Stage 2, Participant C will be able to deduct points from Anna's account and/or Participant B's accounts, depending on the choices they made in Stage 1. In the event that Participant C decides to deduct points, Anna's account and/or that of Participant B will be reduced accordingly.

Specific instructions follow in the next screen.

IMPORTANT: to make sure you understood the decision problems, you will be asked some simple questions, each of which has only one correct answer. If you fail to correctly answer any of those questions, the survey will automatically end and you will not receive any redemption code and consequently you will not get any payment.

With this in mind, do you wish to continue? (Available answers: Continue/End)

*Page break*

Please, read carefully the following instructions.

FIRST STAGE

Anna and Participant B each start this stage with 20 points. There is also a common pool with 0 points at the beginning of this stage. Anna and Participant B can choose to contribute, or not, all of their points to the common pool. No intermediate contributions are allowed. If one contributes 20 points to the common pool, the points are multiplied by 1.5; that is, the amount of points in the common pool will become 20 x 1.5 = 30 points. After Anna and participant B have decided whether or not to contribute to the common pool, the total amount of points in the common pool will be split equally between the two of them, whether they contributed to it or not.

Importantly, Anna and Participant B cannot communicate. They have to make their choice independently of each other and they cannot decide on a joint strategy.

Thus, in summary:

If both Anna and Participant B contribute their 20 points to the common pool, then the total amount of points in the common pool is 30 + 30 = 60 points, which must then be split equally between Anna and Participant B. Thus, at the end of the first stage, Anna and Participant B, will each have 30 points = 20 (starting amount) - 20 (contribution) + 30 (half of the common pool).

If Anna contributes her 20 points to the common pool, but Participant B does not do so, then the total amount of points in the common pool will be 30 points. Thus, at the end of the first stage, Anna will have 15 points (20 - 20 + 15), while Participant B will have 35 points (20 - 0 + 15).

Similarly, if Anna does not contribute her 20 points to the common pool, but participant B does, then at the end of the first stage, Anna will have 35 points while Participant B will have 15 points.

If neither Anna nor Participant B contributes their 20 points to the common pool, then, at the end of the first stage, both Anna and Participant B will have 20 points.

This is the end of the first stage.

SECOND STAGE

The observer, Participant C, who will learn the decisions Anna and Participant B made in stage 1, is given 40 points. Participant C can use these points to deduct points from Anna's account and/or Participant B's account. For every deduction point Participant C uses, the account of Participant C will be reduced by one point while either Anna's or Participant B's accounts will be reduced by three points.

For example, if Participant C decides to use 2 deduction points to deduct points from Anna's account and 5 deduction points to deduct points from Participant B's account, then:

Participant C will end this second stage with 40 - 2 - 5 = 33 points;

Anna will end it with whatever amount she got in the first stage minus 2*3 = 6 points;

and Participant B will end this stage with whatever amount they got in the first stage minus 5*3 = 15 points.

No negative amounts are allowed. Participant C can never deduct more points than the number of points that Anna and Participant B have in their accounts at the end of Stage 1.

At the end of this second stage, the survey ends. Anna, Participant B, and Participant C will be paid according to the number of points they have accumulated during the two stages of the survey (plus the 50 cents for their participation in this HIT). 1 point will correspond to 1 cent.

None of the participants will ever be aware of the identities of the other participants, either during or after this survey, ensuring total anonymity of all persons involved.

Now we ask you some comprehension questions to ascertain that you understood the decision problem.

In Stage 1, what is the choice that Anna should make in order to maximize her number of points? (Available answers: Contribute the 20 points to the common pool/Do not contribute the 20 points to the common pool)

In Stage 1, what is the choice that Anna should make in order to maximize the amount in the COMMON POOL? (Available answers: Contribute the 20 points to the common pool/Do not contribute the 20 points to the common pool)

In Stage 1, what is the choice that Participant B should make in order to maximize Participant B's number of points? (Available answers: Contribute the 20 points to the common pool/Do not contribute the 20 points to the common pool)

In Stage 1, what is the choice that Participant B should make in order to maximize the amount of points in the COMMON POOL? (Available answers: Contribute the 20 points to the common pool/Do not contribute the 20 points to the common pool)

In Stage 2, what is the choice that the observer, Participant C, should make in order to maximize Participant C's number of points? (Available answers: Do not deduct points from anyone/Deduct points from someone)

In Stage 2, what is the role of the observer, Participant C? (Available answers: Deduct points from Anna and/or Participant B, depending on their choices/Add points to Anna and/or Participant C, depending on their choices)

Assume that Anna ended Stage 1 of the HIT with **x** points and that Participant B ended Stage 1 of the HIT with **y** points. If the observer, Participant C, decides to use 2 deduction points to deduct points from Anna's account and decides to use 2 deduction points to deduct points from Participant B's account, how many points will Anna and Participant B each have at the end of Stage 2? (Available answers: Anna gets x - 2 points and Participant B gets y - 2 points/Anna gets x - 6 points and Participant B gets y - 6 points/Anna gets x + 2 points and Participant B gets y + 2 points/Anna gets x + 6 points and Participant B gets y + 6 points)

*Page break*

You passed the comprehension questions.

**You are Participant C**, the observer, and you have 40 points in your account. You have to decide whether or not to use any of your points to deduct points from Anna's and/or Participant B's accounts. You will have to make a decision in each of the following eight scenarios, one of which will be randomly selected at the end of the survey to determine your payment.

*Page break (this and the next three pages are presented in random order)*

Assume that:

Anna **contributed** 20 points to the common pool
Participant B **contributed** 20 points to the common pool
(so, at the end of Stage 1, Anna and Participant B each have 30 points in their accounts)

How many of your 40 points, if any, do you wish to use to deduct three times that number of points from **Anna's** account? (Available answers: slider from 0 to 10 points)

*Page break*

Assume that:

Anna **did not contribute** 20 points to the common pool
Participant B **contributed** 20 points to the common pool
(so, at the end of Stage 1, Anna has 35 points in her account and Participant B has 15 points)

How many of your 40 points, if any, do you wish to use to deduct three times that number of points from **Anna's** account? (Available answers: slider from 0 to 11 points)

*Page break*

Assume that:

Anna **did not contribute** 20 points to the common pool
Participant B **did not contribute** 20 points to the common pool
(so, at the end of Stage 1, Anna and Participant B each have 20 points in their accounts)

How many of your 40 points, if any, do you wish to use to deduct three times that number of points from **Anna's** account? (Available answers: slider from 0 to 6 points)

*Page break*

Assume that:

Anna **contributed** 20 points to the common pool
Participant B **did not contribute** 20 points to the common pool
(so, at the end of Stage 1, Anna has 15 points in her account and Participant B has 35 points)

How many of your 40 points, if any, do you wish to use to deduct three times that number of points from **Anna's** account? (Available answers: slider from 0 to 5 points)

*Page break (this and the next three pages are presented in random order)*

Assume that:

Participant B **contributed** 20 points to the common pool
Anna **contributed** 20 points to the common pool
(so, at the end of Stage 1, Participant B and Anna each have 30 points in their accounts)

How many of your 40 points, if any, do you wish to use to deduct three times that number of points from **Participant B's** account? (Available answers: slider from 0 to 10 points)

*Page break*

Assume that:

Participant B **contributed** 20 points to the common pool
Anna **did not contribute** 20 points to the common pool
(so, at the end of Stage 1, Participant B has 15 points and Anna has 35 points)

How many of your 40 points, if any, do you wish to use to deduct three times that number of points from **Participant B's** account? (Available answers: slider from 0 to 5 points)

*Page break*

Assume that:

Participant B **did not contribute** 20 points to the common pool
Anna **did not contribute** 20 points to the common pool
(so, at the end of Stage 1, Participant B and Anna each have 20 points in their accounts)

How many of your 40 points, if any, do you wish to use to deduct three times that number of points from **Participant B's** account? (Available answers: slider from 0 to 6 points)

*Page break*

Assume that:

Participant B **did not contribute** 20 points to the common pool
Anna **contributed** 20 points to the common pool
(so, at the end of Stage 1, Participant B has 35 points and Anna has 15 points)

How many of your 40 points, if any, do you wish to use to deduct three times that number of points from **Participant B's** account? (Available answers: slider from 0 to 11 points)

**Study 2**

We report the instructions of the *rate-Anna* condition. The instructions of the *rate-Adam* condition are very similar.

*Instructions*

Please, read carefully the following instructions.

Anna and Participant P each start this game with 20 cents. There is also a common pool with 0 cents at the beginning of the game. Anna and Participant P can choose to contribute, or not, all of their cents to the common pool. No intermediate contributions are allowed. If one contributes 20 cents to the common pool, the cents are multiplied by 1.5; that is, the amount of cents in the common pool becomes 20 x 1.5 = 30 cents. After Anna and participant P have decided whether or not to contribute to the common pool, the total amount of money in the common pool will be split equally between the two of them, whether they contributed to it or not.

Importantly, Anna and Participant P cannot communicate. They have to make their choice independently of each other and they cannot decide on a joint strategy.

Thus, in summary:

If both Anna and Participant P contribute their 20 cents to the common pool, then Anna and Participant P will end the game with 30 cents each.

If Anna contributes her 20 cents to the common pool, but Participant P does not, then Anna will end the game with 15 cents, while Participant P will end the game with 35 cents.

Similarly, if Anna does not contribute her 20 cents to the common pool, but participant P does, then Anna will end the game with 35 cents while Participant P will end the game with 15 cents.

If neither Anna nor Participant P contribute their 20 cents to the common pool, then Anna and Participant P will end the game with 20 cents each.

This is the only interaction between Anna and Participant P. They will be paid according to the number of cents they have accumulated.

None of the participants will ever be aware of the identities of the other participants, either during or after this game, ensuring total anonymity of all persons involved.

Now we will ask you several questions, to make sure that you understand the decision problem that the two participants are facing.

Which action should Anna take if she wants to maximise her gain? (Available answers: do not contribute/contribute)

Which action should Anna take if she wants to maximise Participant P's gain? (Available answers: do not contribute/contribute)

Which action should Participant P take if he/she wants to maximise his/her own gain? (Available answers: do not contribute/contribute)

Which action should Participant P take if he/she wants to maximise Anna's gain? (Available answers: do not contribute/contribute)

*Page break*

You passed all comprehension questions. Now please answer the following questions.

*Page break*

Assume that Anna decided to CONTRIBUTE. How would you rate her behaviour? (Available answers: 1 star/2 stars/3 stars/4 stars/5 stars)

Assume that Anna decided NOT TO CONTRIBUTE. How would you rate her behaviour? (Available answers: 1 star/2 stars/3 stars/4 stars/5 stars)

**Study 3a**

The instructions of Study 3a were identical to the instructions of Study 2, apart from two differences:

1. The name "Participant P" was replaced by "Chris".
2. After the rating decision and before the demographic questionnaire, participants were asked two questions:
    a. While answering the questions, what gender did you think Anna/Adam was? (Available answers: Male/Female/Neutral/I didn't know or I didn't think about it)
    b. While answering the questions, what gender did you think Chris was? (Available answers: Male/Female/Neutral/I didn't know or I didn't think about it)

**Study 3b**

These instructions are identical to those of Study 3a, with the only difference that the name "Chris" was replaced by "Skyler".